\documentclass[sn-mathphys,Numbered]{sn-jnl}

\usepackage{graphicx}%
\usepackage{multirow}%
\usepackage{amsmath,amssymb,amsfonts}%
\usepackage{amsthm}%
\usepackage{mathrsfs}%
\usepackage[title]{appendix}%
\usepackage{xcolor}%
\usepackage{textcomp}%
\usepackage{manyfoot}%
\usepackage{booktabs}%
\usepackage{algorithm}%
\usepackage{algorithmicx}%
\usepackage{algpseudocode}%
\usepackage{listings}%
\usepackage{caption}
\usepackage{subcaption}
\usepackage{array}
\usepackage{stfloats}
\usepackage[algo2e,ruled,linesnumbered]{algorithm2e}

\newcommand*{\blue}{\textcolor{black}}

\begin{document}

%

\title{Multi-Objective Application Placement in Fog Computing Using Graph Neural Network-Based Reinforcement Learning}

\author*[1]{\fnm{Isaac} \sur{Lera}}\email{isaac.lera@uib.es}
\equalcont{These authors contributed equally to this work.}

\author[1]{\fnm{Carlos} \sur{Guerrero}}\email{carlos.guerrero@uib.es}
\equalcont{These authors contributed equally to this work.}

\affil[1]{\orgdiv{Department of Computer Science}, \orgname{Universitat de les Illes Balears}, \orgaddress{\street{crt. Valldemossa}, \city{Palma}, \postcode{07122}, \country{Spain}}}



\abstract{We propose a framework designed to tackle a multi-objective optimization challenge related to the placement of applications in fog computing, employing a Deep Reinforcement Learning (DRL) approach. Unlike other optimization techniques, such as Integer Linear Programming or Genetic Algorithms, DRL models are applied in real-time to solve similar problem situations after training.

Our model comprises a learning process featuring a Graph Neural Network and two Actor-Critics, providing a holistic perspective on the priorities concerning interconnected services that constitute an application. The learning model incorporates the relationships between services as a crucial factor in placement decisions: services with higher dependencies take precedence in location selection. 

Our experimental investigation involves illustrative cases where we compare our results with baseline strategies and genetic algorithms. We observed a comparable \blue{Pareto set} with negligible execution times, measured in the order of milliseconds, in contrast to the hours required by alternative approaches.}

\keywords{Fog Computing, Fog Placement Problem, Deep Reinforcement Learning, Graph Neural Network}



\maketitle

\section{Introduction}\label{sec1}

Fog computing extends the cloud paradigm to intermediate or edge devices with computing and storage capabilities to support applications. In this cloud-continuum paradigm, various challenges emerge, encompassing aspects such as security, billing, and the hybridization of ecosystems and providers, among other issues~\cite{Buyya2019FogAE}. In this work, we focus on the decision of choice devices to host applications, known as the Fog Placement Problem (FPP). The FPP represents a combinatorial optimization challenge in placing applications on network devices. Through a well-crafted strategy, we can enhance traditional network metrics like latency and bandwidth, while also considering additional factors such as cost, energy consumption, security, confidentiality, and data resiliency. 
\blue{The FPP is often framed as a bin packing problem, proven to be NP-hard~\cite{NPHard, sami2020dynamic} due to the heterogeneity and large number of devices, the diversity of applications composed of interconnected resources, the variety of placement constraints, and the way previous decisions impact future placements by reducing the available options.}

\blue{Various approaches to designing placement strategies exist in the literature, spanning classical methods based on Integer Linear Programming to those incorporating innovative strategies with evolutionary algorithms or machine learning approaches~\cite{brogi2020place, Salaht2020, fahimullah_machine_2024}}.  \blue{The main disadvantage of ILP is its scalability. While ILP can guarantee globally optimal solutions, it becomes computationally infeasible for large-scale problems due to the exponential growth of the solution space and the complexity of solving large instances. Therefore, evolutionary strategies have gained popularity, with some focusing on optimizing a single objective} such as response time, network bandwidth, economic cost, or energy consumption of placements, while others treat the problem as a multi-objective optimization problem.
\blue{These problems require finding a set of solutions that represent the best trade-offs among the objectives, known as the Pareto front. Unlike single-objective optimization, where one optimal solution is sought, multi-objective optimization results in a set of equally valid solutions, each reflecting different priorities or trade-offs among the objectives. The goal is to identify solutions where no objective can be improved without compromising another~\cite{Li2024, MOOP1}.}
This information proves valuable for decision-makers in determining the optimal balance for the entire set of objectives. For example, placing an application on devices closer to the user may reduce latency but could have higher economic costs compared to placing it in the cloud.

Deep Reinforcement Learning (DRL) is a subfield of reinforcement learning that involves training decision-making between an agent and an environment in dynamically interactive situations. The agent takes an action on the state of the environment and receives a reward or penalty based on the quality of the action. The agent learns a decision-making policy that maximizes the long-run reward~\cite{bookDong}. The term \emph{deep}  refers to the incorporation of neural networks to represent the agent's policy or value function. \blue{DRL has successfully tackled various challenges in domains such as gaming, finance, robotics, and the Fog computing paradigm~\cite{Henderson2018,allaoui_reinforcement_2024}.} and has significantly enhanced the effectiveness of training and exploration within action spaces.


\blue{In our case, the agent determines the deployment location for an application, impacting both response time and the economic cost of each task placement. These are the two objectives we aim to optimize.}
Generally, and particularly in the context of FPP, DRL approaches require extensive training across a great combinatorial space, avoiding the need for testing numerous heuristics~\cite{bookDong}. Once trained, they can swiftly provide solutions to new cases. In the relevant literature, various studies explore different DRL-based learning modelling techniques, as documented by Goudarzi et al.~\cite{Goudarzi2021}. Their article delves into the analysis of criteria considered in other related works for the agent's learning policy, such as time and energy consumption.

Another contribution of our work involves the agent learning about the priorities of tasks within an application. Essentially, the deployment of an application can be viewed as the deployment of a set of services/tasks. Rather than making a single placement decision for an application, multiple decisions can be made from this second perspective. It aligns the design of applications with paradigms such as microservices architectures, making the placement problem a more fine-grained problem. The priority of services/tasks is determined by the communication dependencies among other services/tasks and their processing time. Placing higher priority tasks on appropriate devices and assigning the remaining tasks to those devices or ones closer in terms of computational capabilities and network latencies results in the application offering the user a lower response time. Conversely, scattering higher priority tasks on devices where their communication dependencies are not aligned in the infrastructure, will lead to more network traffic and a worse response time. To the best of our knowledge, only Wang et al.~\cite{Wang2020FastAT} and Goudarzi et al.~\cite{Goudarzi2021} have incorporated task priority into their approaches in a more straightforward manner.

In our study, we model the dependency graph among tasks into a specific family of graph neural networks called Graph Isomorphism Networks (GIN)~\cite{Xu2018HowPA}. In the GIN model, the features of tasks include the execution times and their dependencies. GINs facilitate the propagation of features between neighbouring nodes over multiple iterations, aiding the identification and classification of nodes. The classification of each task provides a score that determines the priority in our DRL model. 
Additionally, the agent's policy is based on the use of Proximal Policy Optimization (PPO)~\cite{Schulman2017ProximalPO} with two agent-critic models. The first actor-critic in our model considers the environment state in terms of task features and dependencies from the GIN model and selects a task. The second actor-critic uses the current task placement, the state of the devices, and the selected task to allocate it to one of these devices. 

Another aspect to consider in the use of DRL in a multi-objective problem is that they typically provide a single solution driven by the reward values received by the agent during its learning process~\cite{6918520}.
However, in multi-objective problems, it is common to obtain numerous solutions that are valid based on the desired balance across each objective. \blue{The ability to generate a solution set using DRL is particularly motivating because DRL approaches typically provide responses based on the current input state. This is a fundamental characteristic of DRL algorithms, which operate within a continuous cycle of observation and action.} To achieve this spectrum of solutions, known as the \blue{Pareto set}, we perform scalar decomposition, necessitating the generation of different learning models according to the desired balance of objectives. This balance imparts a different guide to rewards. For instance, an agent under a model prioritizing latency will exhibit completely different behaviour than a model that simultaneously balances latency and energy consumption. Given that training each model is computationally expensive, we mitigate costs by transferring parameters between models. This approach allows us to quickly identify dominant solutions in the problem space and delegate decision-making to the manager~\cite{goldberg1989genetic}.

The key contributions of this paper can be summarized as follows: (I) the utilization of a GIN for handling priority placements based on the dependencies among the services constituting an application, offering a more comprehensive and holistic perspective on service selection, and (II) the generation and analysis of a multi-objective front through parameter transfer between models.

\blue{The structure of this article comprises four main sections. The first section offers an analysis of the state of the art, exploring various strategies based on neural networks to address the FPP. In the second and third sections, we elaborate on the adopted model, the DRL-agent environment, and the neural models. In the final section, we present the experiments conducted to evaluate our proposal in terms of the range of solutions, training times, and applicability time. We compare our approach with simple strategies and two types of genetic algorithms: a single-objective and a multi-objective approach, specifically NSGA-II.}

\section{Related work}

The application of reinforcement learning algorithms to optimization problems has been extensively researched, including within the domain of cloud computing~\cite{arxiv.2105.04086}. In the Fog computing domain, the use of deep learning techniques has opened up new research opportunities, covering a broad spectrum of optimization cases. These cases include task offloading, scheduling, resource allocation, load balancing, resource discovery, and content caching~\cite{IFTIKHAR2023100674}.

Concerning our proposal, we primarily focus on placement, scheduling, and offloading problems because all of them are related to our objective of selecting and placing services in Fog environments. In this context, related studies explore various reinforcement learning algorithms or techniques, such as Q-Learning~\cite{farhat2020reinforcement, Nassar2019}, Deep Q-Networks (DQN)~\cite{Zheng2022Too, Mseddi2019, li2020intelligent, LiHe2019, Poltronieri2021ReinforcementLF}, model-agnostic meta-learning (MAML)~\cite{Wang2020FastAT}, or variants of actor-critic algorithms like actor-critic based approaches~\cite{SACC22}, importance-weighted actor-learner architectures (IMPALA)~\cite{Goudarzi2021}, and Proximal Policy Optimization (PPO)~\cite{PPO2022}. These techniques often differ in policy update, training stability, degree of exploration, or hyperparameter tuning. In our proposal, we use an actor-critic model with PPO because it has demonstrated good stability of results with a relatively low number of training episodes and few hyperparameter settings. In our experiments, the training range is significantly lower compared to other studies, which require hundreds or even thousands of episodes for training.

The related proposals optimize various performance indicators, such as service time, the response time (including communication times), or the reduction of data transfer in offloading problems. For example, Farhat et al.~\cite{farhat2020reinforcement} propose an approach to predict the time and location of required fogs, aiming to minimize the number of requests processed by the cloud. \blue{In other words, they aim to reduce the cloud's load by maximizing the use of available fog resources across different locations. Our multi-objective model takes both time and cost into account; while load is not a primary objective, it can be incorporated into the reward calculation to guide the learning process, as will be discussed later.} Zheng et al.~\cite{Zheng2022Too} focus on time optimization, considering parameters like task properties, network situation, and computational resources to make offloading decisions. \blue{In their case, there are no dependencies between the tasks.}  Mseddi et al.~\cite{Mseddi2019} aim to improve the number of satisfied user requests within a predefined delay threshold. \blue{They use a network model called a deep Q-network (DQN) to approximate the Q-function.} In the case of Li et al.~\cite{li2020intelligent}, the learning model's goal is to reduce the workload distributed among devices. \blue{An interesting contribution of their work is the definition of the reward, which considers the weighted sum of the remaining time to complete the execution of a task, the energy consumption of the user device, and the server load}. Zhou et al.~\cite{SACC22} propose a cache-enhancement system where the hit rate serves as the actor's reward. In our case, our goal is the overall execution time of applications, encompassing both the service time and the communication time between dependent services. \blue{Zhang et al.~\cite{zhang_drl-based_2024} address computation offloading in high-mobility Internet of Vehicles environments, focusing on latency, energy consumption, and cost challenges. The proposed decentralized framework leverages Multi-Agent Deep Reinforcement Learning to optimize task-resource matching, overcoming the limitations of traditional DRL approaches that treat agents independently.} \blue{Bai et al.~\cite{bai2021} propose a DRL-based method for optimizing offloading strategies and resource allocation in fog computing to minimize system costs. By using a multi-agent setup, it overcomes the dimensional explosion problem of traditional DRL. Experiments show that the proposed method effectively reduces costs compared to random strategies across different numbers of fog nodes and users.}

About the management of the task priority, Goudarzi et al.~\cite{Goudarzi2021} propose a method for scoring services in a pre-scheduling phase, which generates an ordered sequence of services based on the weighted average cost of executing the task on different infrastructure devices. Then, the sorted selection of services in this sequence determines the service placement. Additionally, this sequence provides information about the ranking of parallel services by giving higher priority to tasks in this subset of parallelisation with a longer execution time. Wang et al.~\cite{Wang2020FastAT} also generate task sequences, but in their case, it is for offloading decisions to reduce latency. In our approach, we propose that task selection should be another decision in the learning model. Thus, we avoid a previous phase of scheduling. We use a graph neural network model that takes as input the cost of operations for each task and the adjacency matrix, as we explain in the next section. 



\section{Problem and system model}

In this section, we describe the problem and our framework. Since there are numerous terms, we provide the most important in Table~\ref{tab:listaanotaciones} in advance. 

Our fog infrastructure consists of a collection of devices, each with a unique identifier \blue{($k$) and three features: speed of operation execution  ($Speed_k$), latency ($Lat_k$) and cost ($Cost_k$)}. Devices with low latencies mean that they are closer to each other. \blue{The cost is related to the device execution speed; it is a normalized value within the speed range, where higher $Speed_k$ incurs higher $Cost_k$.} In the infrastructure, one device represents the cloud entity. 

The application model commonly used in the literature places a single application in the fog environment or considers it a set of services with dependencies represented by a Directed Acyclic Graph (DAG)~\cite{8588297}. To standardize the representation of applications in our model, we adopt the same representation of operations as in the Job-Shop Scheduling Problem (JSSP), which is also based on a DAG~\cite{Applegate1991ACS}. 
\blue{A JSSP instance consists of a sequence of ordered operations, also known as tasks}.  In our approach, these operations correspond to the services/tasks defining the application, and the order between operations signifies the dependency relationship between the services. The size of an application is proportional to the number of services, so an instance of size 3 contains $3\times 3 = 9$ services. Each row of the matrix represents an ordered sequence of communications between services: $S_{i1}\rightarrow S_{i2}\rightarrow ... \rightarrow S_{in}$. With this modelling approach, it is possible to represent application models that do not adhere to the structure of $N\times N$ services. \blue{We can easily translate this matrix structure representing the dependencies of an application's services into the input for a neural network, as will be demonstrated in the following sections.}

\begin{table}[!t]
\caption{List of mathematical symbols.\label{tab:listaanotaciones}}
\centering
\begin{tabular}{|l|p{6.5cm}|}
\hline
$T_{app}$ & Response time of a deployed application \\\hline
$S_{ij}$ & \blue{Service located in the i-th row and j-th column of the dependency matrix within the application} \\\hline
$S_{i1}\rightarrow S_{i2} $ & Service dependency\\\hline

$O_{i,j}$& Number of operations of the service related by position\\\hline
$k$& K-th device \\\hline
$Speed_k$& Device execution speed\\\hline
$Cost_k$& Device cost\\\hline
$Lat_k$& Device communication latency \\\hline
$f_i$ & Optimization objective functions\\\hline
$feat_{devices}$ & Device features\\\hline
$feat_{service}$ & Service features\\\hline
$t $ & \blue{The temporal index organises the sequence of actions, states, and rewards}\\\hline
$a_t$ & Action, a placement decision\\\hline
$s_t$ & State \\\hline
$r_{f_i}$ &  Specific reward for each optimization objective function\\\hline
$r_t$ & Total reward of an action\\\hline
$\pi$ & Policy \\\hline
\blue{$AC_s$} & Actor-Critic for Service selection\\\hline
\blue{$AC_d$} & Actor-Critic for Device selection\\\hline
$ \theta_{AC_i}$ & Weights and biases of the model \\\hline
\end{tabular}
\end{table}

\begin{figure}[!t]
\centering
    \includegraphics[width=2.in]{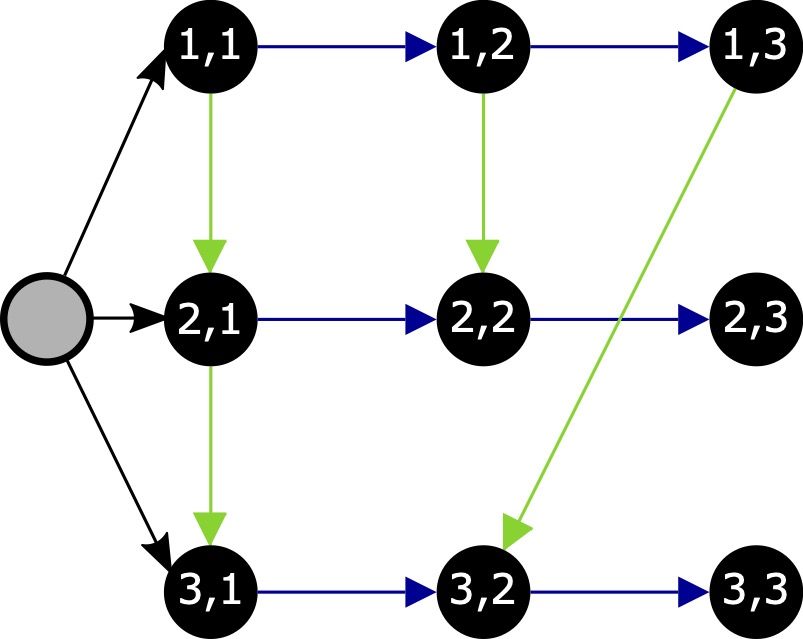}
    \hspace{0.5cm}
    \includegraphics[width=2.in]{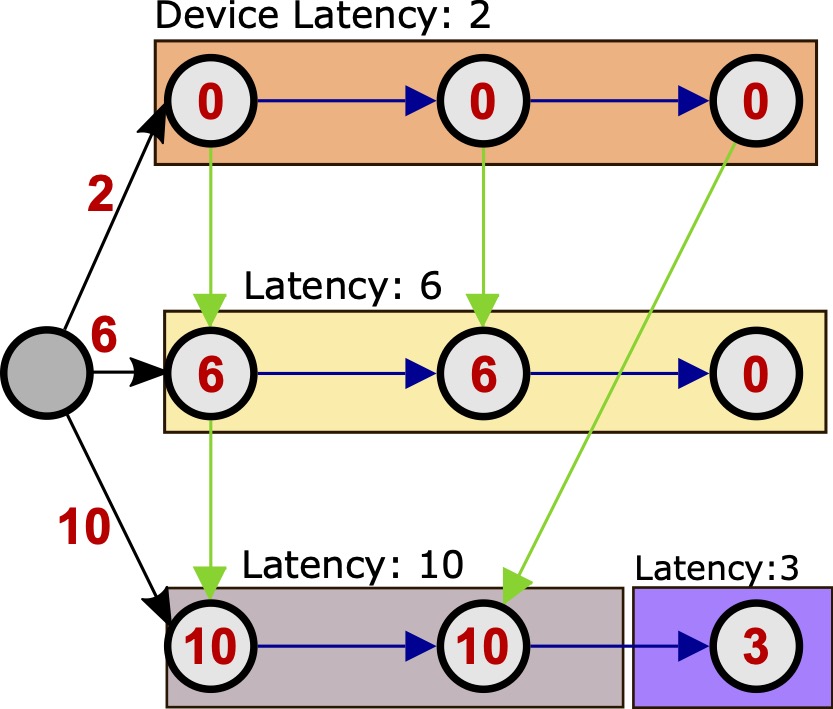}%
\caption{Example of an application based on a JSSP instance with $3\times3$ services. The left figure contains the communication dependencies. \blue{The right figure illustrates the assignment to specific devices (represented by squares) and their corresponding latency values for a calculation scenario.}}
\label{fig:appExample}
\end{figure}

Figure~\ref{fig:appExample} (left figure) represents an example of an instance of size 3 with $3\times 3$ services. The grey node is an abstract service that represents the whole application through the composition of its services. In addition to the three rows of common dependencies, we generate extra dependencies between predecessors' services. In this example, there are 4 more: $S_{11}\rightarrow S_{21}$, $S_{12} \rightarrow S_{22}$, $S_{21} \rightarrow S_{31}$ and $S_{13} \rightarrow S_{32}$. Each service $s_{ij}$ has an execution time that we calculate by the number of operations ($O_{i,j}$) divided by the execution speed ($Speed_k$) of the device deployed ($k$). We use abstract units to simplify calculations and enhance readability. A device also has two other features: a cost $Cost_k$ and a communication latency $Lat_k$. 

To enhance clarity, we decompose the computation of the application response time, denoted as $T_{app}$, into three equations. Equation~\ref{eq:tiempoT} calculates the execution time of each service by dividing the number of operations by the CPU speed of the device. The device assignment is a binary indicator variable $s_{i,j}^k$, where the value is 1 if the service $s_{i,j}$ is assigned to device $k$, and 0 otherwise. Equation~\ref{eq:tiempo2} calculates the access latencies to the initial services of the application. Equation~\ref{eq:tiempo3} calculates the latencies between dependencies of services that are hosted on different devices $[k \ne q]$. 
When all services are on the same device we only consider the latencies of the starting services of each row (\(s_{i,1}\)). The other dependencies do not influence the communication cost since all the services are in the same entity. For example, we consider that the services are initially on the cloud device. In this case, we achieve a reasonably intermediate response time, as observed in the experiments, since all services are together. 

\begin{align}\label{eq:tiempoT}
T_{app} = \sum_{i=1}^{n} \sum_{j=1}^{n} s_{i,j}^k \cdot \left( O_{i,j}/Speed_k  \right) +\\\label{eq:tiempo2}
+ \sum_{i=1}^{n} s_{i,1}^k \cdot Lat_k  + \\\label{eq:tiempo3}
+ \sum_{i=1}^{n} \sum_{j=1}^{n} \sum_{a=1}^{i} \sum_{b=1}^{j} \left( s_{i,j}^k \rightarrow s_{a,b}^q \right) [k \ne  q]\cdot Lat_q 
\end{align}

To illustrate the calculation process of \(T_{\text{app}}\) based on one placement case, we consider the application depicted in Figure~\ref{fig:appExample} (right figure). It contains the placements of the nine services on four devices with different latency values. \blue{These placements represent the decision vector for our algorithm: $S_{1,1}$ on $D_{orange}$, $S_{1,2}$ on $D_{orange}$, $S_{1,3}$ on $D_{orange}$, $S_{2,1}$ on $D_{yellow}$,\ldots,$S_{3,3}$ on $D_{purple}$.} In this example, we make the following assumptions: (i) To simplify the calculation, we consider the execution time of the nine services as negligible, resulting in \(O_{i,j}/\text{Speed}_k\) being zero; (ii) services in the same row are deployed on the same machine with latencies 2, 6, and 10, respectively; (iii) service \(S_{3,3}\) is deployed on a device with latency 3. Thus, the $T_{\text{app}}$ is 53, which is the sum of the access latencies to the first service of each row (i.e., 2, 6, and 10) and the latencies for communication dependencies between services, which we represent as a matrix: [[0, 0, 0], [6, 6, 0], [10, 10, 3]]. \\


Our objective is to optimize the response time and cost of the services that compose an application.  The formalization of the multi-objective problem is as follows: 

\begin{align}
\label{eq:goals1} 
\min_{f_1} \quad & f_1 = T_{app} \\
\label{eq:goals2}
\min_{f_2} \quad & f_2 =  \sum_{i=1}^{n}\sum_{j=1}^{n}s_{i,j}^k \cdot Cost_k \\
\label{eq:goals3}
\textrm{s.t.}  \quad & \sum_{i=1}^{n}\sum_{j=1}^{n} s_{i,j}^k = 2n 
\end{align}

\begin{figure}[!t]
\centering
\includegraphics[width=2.5in]{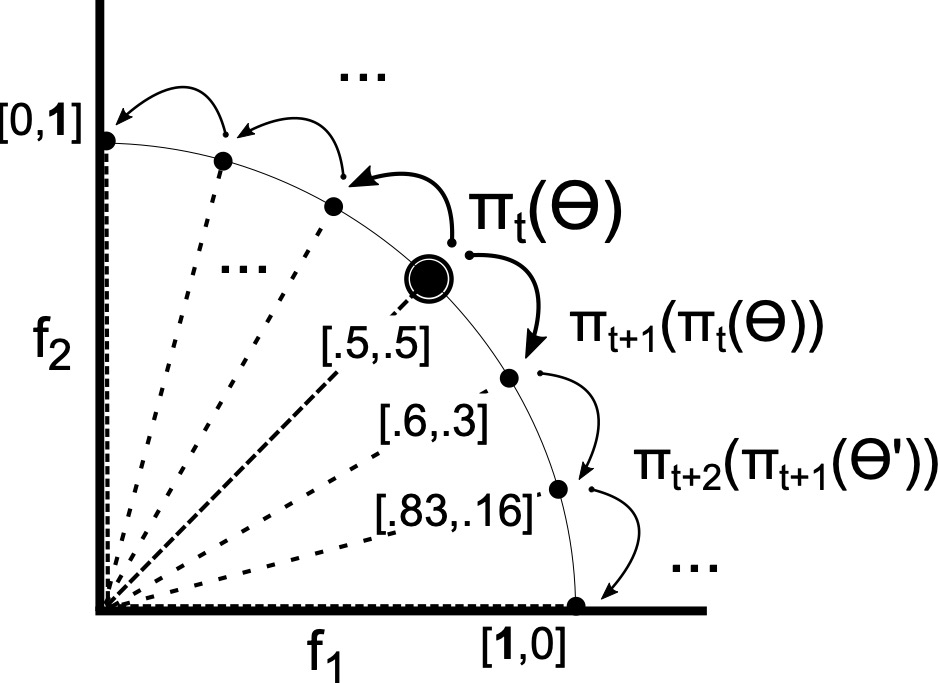}
\caption{Scalar decomposition of objectives based on weights with parameter transfer between models.}
\label{fig:espacioescalar}
\end{figure}

In most cases, solving multi-objective problems involves obtaining a set of non-dominant solutions known as the Pareto set. The Pareto set represents a collection of efficient choices for various parameters, which allows for trade-offs within this set.
To obtain a Pareto-front from a DRL algorithm, we need to drive the rewards of the agent using a weighted average of the goals. Thus, we propose a scalar decomposition approach to obtain this front of solutions~\cite{LiZW21}. Furthermore, we implement a parameter transfer process between models that significantly reduces the cost of the training process for each combination of weights. \blue{We transfer the weights and biases between models. This means that the new model does not need to reinitialize these values from scratch, which significantly reduces the number of training iterations required}. Figure~\ref{fig:espacioescalar} illustrates the parameter transfer strategy used in this study, where we compute the middle model first and parallelize both closest models.

\section{DRL framework}

As mentioned, our goal is to minimize the response time and cost of the application by optimizing the allocation of services. For that, the agent must consider two decisions: selecting the service according to the DAG dependencies of the application and choosing the host device. The agent assigns services individually, taking a holistic approach using the DAG, rather than deploying all services simultaneously. This strategy offers greater flexibility, enabling us to scale high-demand services in production environments, for example, by invoking the agent's policy when new services require scaling. 

\begin{figure*}[!t] 
\centering
\includegraphics[width=1.\textwidth]{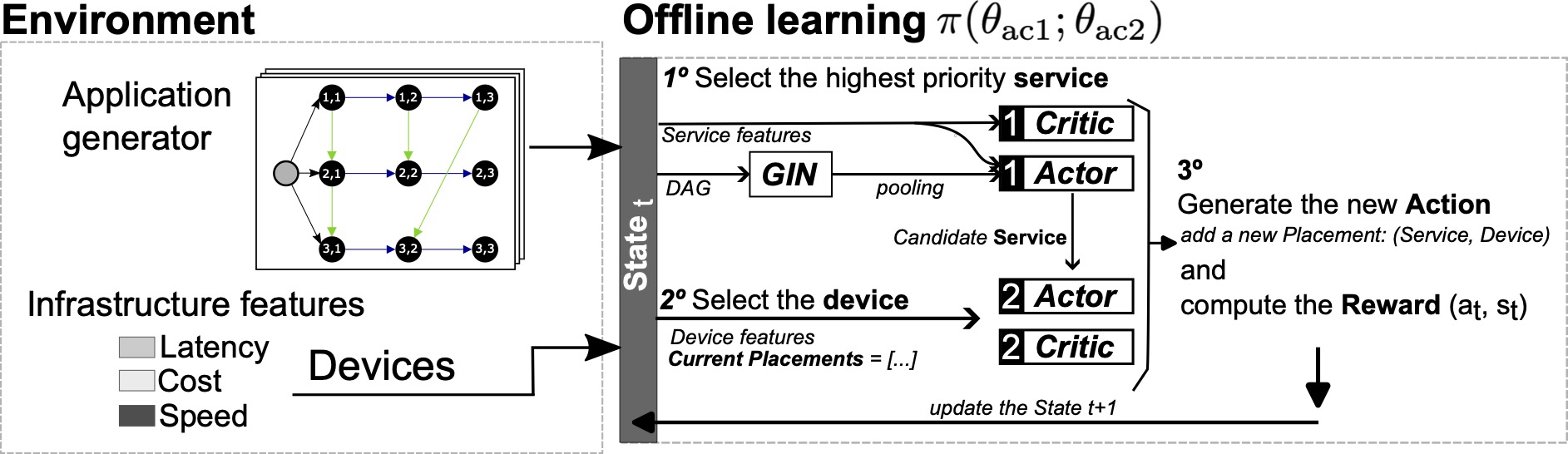}
\caption{Outline of the placement process from the training perspective.}
\label{fig:modeloEsquema}
\end{figure*}

Figure~\ref{fig:modeloEsquema} synthesizes the entities that constitute the training process. The agent learns in an environment set up by devices and applications to deploy with their respective services. This situation gives rise to a state that is the agent's input. The agent's policy uses the DAG and the services' features to make the first decision about the election of service among those that have yet to be deployed. The second decision is the choice of the device for the candidate service. Both decisions make up the service placement action. Each action generates a reward and a new state of the environment, which will be the agent's next input. Periodically, batches of actions refine the agent's learning policy.

\blue{Once the training is completed, we can deploy new applications using the optimal policy obtained. Algorithm~\ref{alg:deplacement} outlines the steps of this process, which is ready to be invoked at any time without requiring additional phases.}
In lines 1-3, we define the system's state concerning the devices, dependencies among application services, and service times. At that point, an iterative process begins to place all the services. The selection policy ($\pi(\theta_{\text{ac1}})$) selects the service, as indicated in line 6, and the second trained policy ($\pi(\theta_{\text{ac2}})$) \blue{chooses} the device. The placement is recorded, and the device's state is updated. The process continues until all services are placed.

\begin{algorithm2e}
\caption{\blue{Optimally trained placement algorithm, prepared for immediate use}}


\label{alg:deplacement}

\KwData{$application, devices$}
\KwResult{$placements$}
devicesSt $\gets$ getStatus(devices)\;
DAG $\gets$ getDAG(application)\;
servicesSt $\gets$ getTimes(application)\;
placements $\gets []$\;
\While{$\exists$ application.services not allocated}{
   service $\gets \pi(\theta_{\text{$AC_t$}}(servicesSt,DAG))$\;
   device $\gets \pi(\theta_{\text{$AC_m$}}(servicesSt,service,devicesSt))$\;
   devices[device].place(service)\; 
   devicesSt $\gets$ updateStatus(devices)\;
   placements.append([service,device])\;
}
\end{algorithm2e}

\subsection{Reinforcement learning environment}

In this section, we describe the state, actions, reward, and policy that are the main components of a standard reinforcement problem.

\subsubsection{State} 
A state represents the situation of the environment. In our model, the state comprises two parts: the completion time of the services/tasks that contains the execution time based on device assignment and any potential latencies from dependencies, and the state of the devices that encompasses the features of the devices and the current task placements on them. Given that we are considering an initial placement, we allocated all the services to the cloud device. In terms of response time, the initial state is a relatively good placement since all the services of an application are on the same device, as only the latencies for each initial service $S_{i1}$ (from the JSSP model we mentioned earlier) need to be considered in that situation. If the cloud entity has both the lowest latency and good execution time, it would be one of the best possible solutions.
The dimension of each part of the state is $Tasks \times Features_{Tasks}$ and $Features_{Devices} \times (Tasks \times Features_{Tasks})$,   where $Features_{Tasks}$ represents the task characteristics, such as the number of operations, and $Features_{Devices}$ denotes the device features, including latency, speed, and cost.

\subsubsection{Action} 
The agent chooses both the task and the device. As the agent's trajectory advances, the count of pending tasks decreases, resulting in a reduction of the agent's decisions. In both, the action space is discrete. The number of tasks determines the quantity of actions taken and, consequently, the length of the agent's trajectory.

\subsubsection{Reward} 
The reward value calculated for each placement action is crucial for aligning the objectives ($f_i$) presented in equations~\ref{eq:goals1} and \ref{eq:goals2} with the learning phase. The reward is calculated based on the qualitative difference in response time and cost gain between the two states: $r_{f_i}(a_t, s_t) = r_{f_i}(s_t) - r_{f_i}(s_{t+1})$, where $r_{f_i}(\cdot)$ represents the quality function for the objective. We integrate both individual rewards through a function that normalizes and balances both objectives, expressed by the following equation:

\begin{equation}
   r = \sum_{i=1}^k w_i \cdot r_{f_i}(\cdot)
\end{equation}

Here, \(w_i\) represents the weighting value ranging from 0 to 1, and \(r_{f_i}(\cdot)\) denotes the independent reward of each normalized objective function. Normalization is achieved by range, taking into account that we determine the maximum value through prior experimentation for each specific case.

When the agent finishes the placement of all services, the cumulative sum of the reward gives the application response time under the last assignment made. Maximizing each difference minimizes the optimization objective. Table~\ref{tab:responseexample} contains an example of reward evolution regarding the response time. In this example, 
the initial placement (\textit{Step 0}) of all services is the cloud device, resulting in a response time of 60 units. The two successive actions worsen this time due to the dependencies between the services. The last placement decision (\textit{Step 3}) brings the third service closer to the other two, which results in a positive reward for this action and finally improves the execution time in 47 units compared to its initial location. 

\begin{table}[!t]
\caption{Example of the evolution of the reward over the response time of an application composed of three services.\label{tab:responseexample}}
\centering
\begin{tabular}{|c|r|r|r|r|}
\hline
\textbf{Step} & \textbf{0} & \textbf{1} & \textbf{2 }& \textbf{3}\\\hline\hline
\textbf{Reward} & -60 & -15 & -2 & +30\\\hline
\textbf{$T_{app}$}$^{\mathrm{a}}$ & 60 & 75 & 77 & \emph{47}\\
\hline
\multicolumn{5}{l}{$^{\mathrm{a}}$It is the absolute value of accumulative reward}
\end{tabular}
\end{table}

\subsection{Learning policy}

Our agent explores the environment for a trajectory based on a Stochastic Policy ($\pi(a_i|s_t)$). We train the model using a gradient based on a sequence of weights between a Graph Neural Network (GNN) model and an Actor-Critic (AC) model to identify the task selection action and a second Actor-Critic model to determine the selection of the device.

Both agents use a Proximal Policy Optimization (PPO) to limit the learning value, which results in more stable learning. Additionally,  we train the weights of both actor-critical models with an average combination of both loss values.

\subsubsection{Graph Neural Network (GNN)}

Our approach relies on GNN models to extract aggregate indicators from graphs of variable size. These indicators assess the services based on their dependencies. We utilise a type of GNN known as Graph Isomorphism Networks (GIN) to capture the global dependencies of a service with its neighbours.

GIN operates on the principle of incorporating information from the local neighbourhood of each node in a graph, as opposed to considering individual nodes in isolation. This is achieved through the aggregation of information from neighbouring nodes using a permutation-invariant aggregation function. Xu et al.~\cite{Xu2018HowPA} proposed Multi-Layer Perceptrons (MLPs) for feature extraction in different k-iterations through equations \eqref{eq:gin} and \eqref{eq:gin2}.

\begin{equation}
h^{(k)}_{v} = MLP^{(k)}(
(1+\epsilon^{(k)}) \cdot  h_v^{(k-1)} + \sum_{u\in N(v)} h_u^{(k-1)})\label{eq:gin}
\end{equation}

and 

\begin{equation}
h^{(0)}_{v} = MLP^{(0)}(\textbf x_v )\label{eq:gin2}
\end{equation}

In the initial iteration $k=0$, each node $v$ obtains an MLP-based representation of its features $\textbf x_v$. In successive iterations, the node representation incorporates the aggregation values of its neighbours ($N(v)$) with the previous values. This process is repeated $k$ iterations, with a learning rate $\epsilon$ controlling the step size, and applying a batch normalization to further stabilize the learning. The final representation of the whole graph is obtained by applying an average pooling operation over the node representations: $ h_g=1/|V| sum_v \in V h^K_v $.

The GIN input is the adjacency matrix of the services, the aggregation values of the adjacency matrix and the state representation of the current situation of tasks. The output of the network is the weights of the MLP model that incorporate the state weighting by the aggregation $h_g(s_t)$ and the graph embedding after $k$-iterations $h^{(k)}(s_t)$.

\subsubsection{Actor-Critic networks}

Each Actor-Critic decides by sampling an action from a probability distribution obtained by an MLP model. The first actor-critic chooses a candidate service from those that have already resolved their dependencies, based on a score obtained from the GIN model and a softmax function: $softmax(MLP(h_g(s_t),h^{(k)}(s_t)))$. The second actor-critic selects a device from the available ones, based on the score of the output of its $MLP$ model. The input of the model includes the candidate services, the state of the devices and the current placement, described in eq. \eqref{eq:ac2}. We record both actions and corresponding probabilities for the training process.

\begin{equation}
softmax( MLP (
Features_{Devices}\times Features_{Candidate},Features_{allocation}(s_t)) ) \label{eq:ac2}
\end{equation}

We use the PPO algorithm~\cite{Schulman2017ProximalPO} to implement our two agents. The significant difference is that we use the experience recorded by multiple actors in different environments to perform the update.

\section{Experiments}

We divide this section into four parts: model configuration, where we provide details on the hyperparameters considered; environment configuration of each experiment; evolutive algorithm configuration, where we detail the parameters of both \blue{genetic algorithms}; and finally, we present the results. 

\blue{The use of two genetic algorithms, one multi-objective and the other single-objective, aims to establish a comparative framework for their solutions against our proposed method. The multi-objective algorithm will provide a Pareto front that illustrates the direction, diversity, and density of the solutions. Meanwhile, the single-objective algorithm will be adapted to provide a multi-objective framework, similar to our approach.}

The data generated in this study and the source code of the scripts are publicly available in a repository\footnote{\url{https://github.com/acsicuib/DRL-AC-Allocation}}.

\subsection{Model configuration}

The hyperparameters of the model are the same in all experiments. We explore different hyperparameters to find the most suitable choice. The training lasts for 150 episodes, each of which contains 40 independent environments, or what are the same, 40 agents or trajectories. In all cases, we normalize all features to the same scale.

The GIN model and the service selector $Critic_s$ have an MLP network with an input dimension of $hg(s_t)$ and four layers. Both actors $Actor_s$ and $Actor_d$ are MLP models with five hidden layers, and both $Critic$ networks are also MLP models with three hidden layers. The input shape of $AC_s$ has a dimension of 64, and the input shape of $AC_d$ has the  $Features_{Devices}+NT^2$ size. GIN model plus $AC_s$ and $AC_d$ models are trained by Adam optimizer with a learning rate of 0.022. The gradient ($\epsilon_{ppo}$) is clipped to 0.25. The scheduler StepLR is with a decay rate ($\gamma$) of 0.9. The epochs of updating the network, the policy loss, value function and entropy are set to 2, 3, 2, and 0.023, respectively. Other parameters use the configuration of Pytorch library~\cite{Paszke2019PyTorchAI}. 

\subsection{Experiment configuration}

We conduct two classes of experiments based on the number of jobs, devices and study objectives. We run all the experiments with an Intel(R) Xeon(R) Silver 4214 CPU@ 2.20GHz with 32GB RAM and two Nvidia Quadro RTX 6000 graphic cards.

In the first experiment, we examine the evolution of rewards using various comparative strategies, such as assigning to devices with low cost or low latency, deploying only in the cloud, or doing a random placement. The environment contains 1000 devices and 9 tasks by row (81 services/tasks per application).  

In the second experiment, we assess the scalability of the model and the results. As a baseline, we use a well-known multi-objective genetic algorithm, NSGA-II~\cite{nsga2}, and a single-objective genetic algorithm (GA).  The configuration of the environment spans 500, 1000, 1500, and 2000 devices, with a task range of 9 or 12 per row (81 services or 144 services in total per application, respectively).

We set the number of operations for each task and the device speed to a constant value. This simplifies result interpretation by focusing solely on the latency characteristics in execution time calculation. In addition to the existing task dependencies in the Job-Shop model, there is a 20\% chance that a task has an extra dependency on a predecessor task. The latencies of devices are generated uniformly by selecting one of the values from [1, 10, 20, 30, 40, 50], and its cost is determined based on the values [1, 10, 20, 30, 40]. The cloud device maintains the same features in all scenarios, with a maximum latency of 50 units and a cost of 20 units.

\subsection{Genetic algorithm configurations}

In the second experiment, we use a single-objective \blue{Genetic Algorithm (GA)} and a multi-objective algorithm, NSGA-II. In both, we use the interface provided by the library~\cite{pymoo} to implement both algorithms, which are available in the project's repository.

\blue{Following several tests derived from previous related works where we used a similar model of a Fog ecosystem and applications~\cite{sergiGA}}, we establish a configuration for the 8 test combinations conducted in this second experiment. Hence, we need 200 generations, which yielded acceptable convergence when there were 12 tasks per row, an initial population of 200 solutions, and a 15\% probability of mutation in the population crossover. 

\subsection{Results}


\subsubsection{Experiment I - Comparison of baseline strategies}

We compare the reward achieved by our agent's decisions with other control strategies and examine the stability of the results across different experiment dimensions, as discussed earlier.

The control strategies are (I) a naive strategy that samples random devices, which we named \texttt{Random devices}; (II) a  conservative strategy that keeps all services in the cloud, which we named \texttt{All-in-Cloud}; (III) and finally, we implement two strategies that, according to the experimental setup, should have the same behaviour. These two strategies are both greedy algorithms: the first one \texttt{Greedy-Edge} chooses the device with the lowest priority and among them always chooses the one with the lowest cost; the second one called \texttt{Greedy-Cost} is the opposite of this one, giving initial priority to devices with the lowest price and then chooses the one with the lowest latency.

\begin{figure*}[!t]
\centering
\includegraphics[width=5.0in]{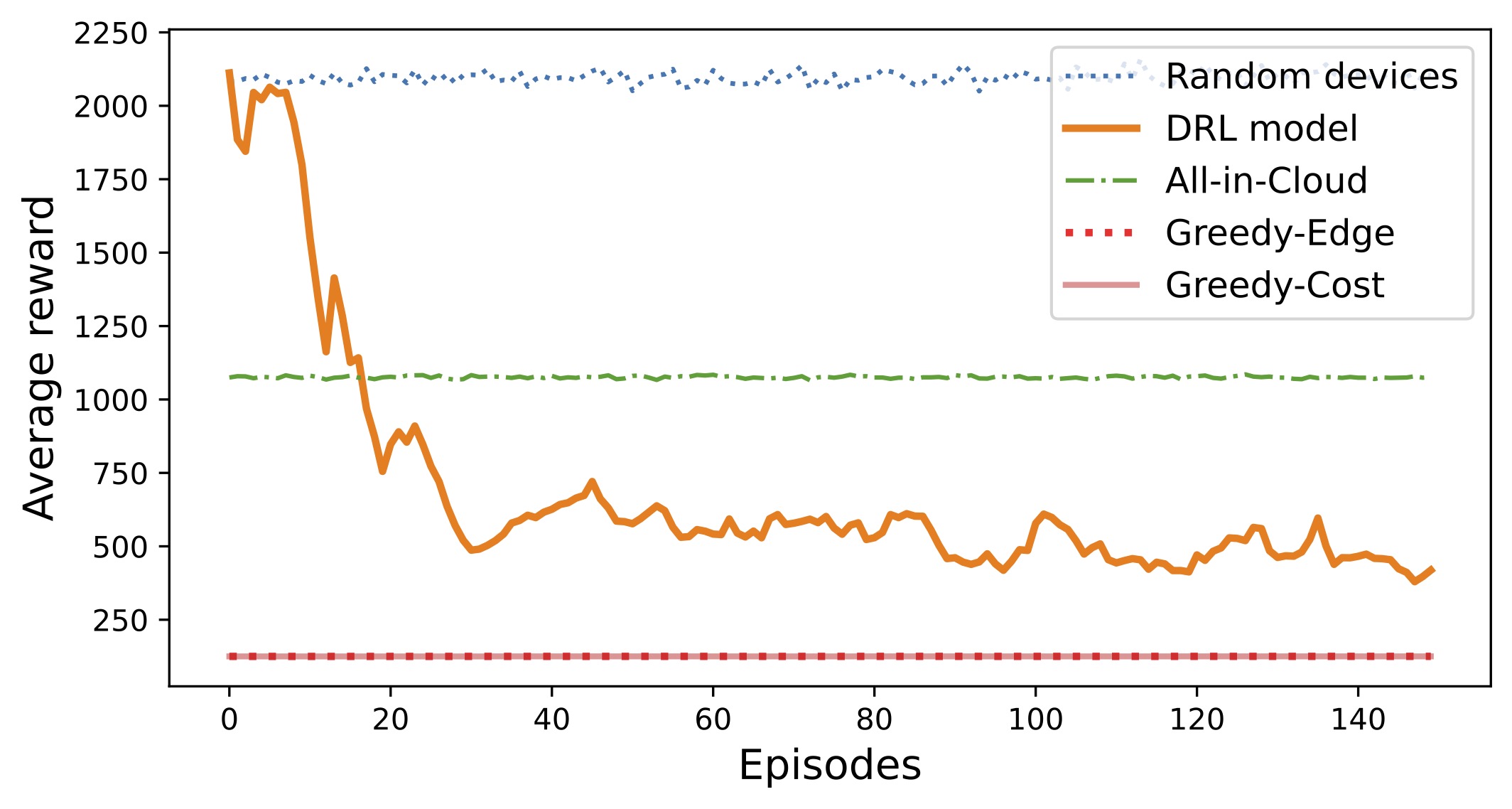}
\caption{Evolution of the reward obtained in a configuration with 1000 devices and 81 tasks with our DRL strategy and different comparative strategies.}
\label{fig:caseII}
\end{figure*}

In Figure~\ref{fig:caseII}, we present the evolution of the reward for each environment or episode of the previous baseline strategies, as well as the behaviour of our trained policy, which we refer to as \texttt{DRL model}. We base the results on a configuration with 1000 devices and 9 tasks per row (81 tasks) with balanced weighted goals, where latency and cost are both at 50\%. Each strategy exhibits an expected behaviour: (I) the random placement strategy (\texttt{Random devices}) is the worst option as it disregards any criteria; (II) the \texttt{All-in-Cloud} strategy maintains the initial placement configuration of services, and since all the services are on the same node, the execution time is not bad, as we will see later;  (III) \texttt{Greedies approaches}, the scenario configuration for this study is set up so that there is at least one device among 1000 randomly generated that has the lowest latency and cost values. This design enables us to position the optimal solution in the solution space compared to other algorithms using a greedy approach. Both greedy strategies obtain the same reward value. The evolution of the reward in our strategy is consistent with the agent's behaviour, as it gradually improves its choice of location as it learns. During the first episodes, it exhibits exploratory behaviour similar to the random strategy. Around episode 20, it reaches the cloud placement results and stabilizes around episode 60.

\begin{figure*}[!t]
\centering
\subfloat[]{\includegraphics[width=5.0in]{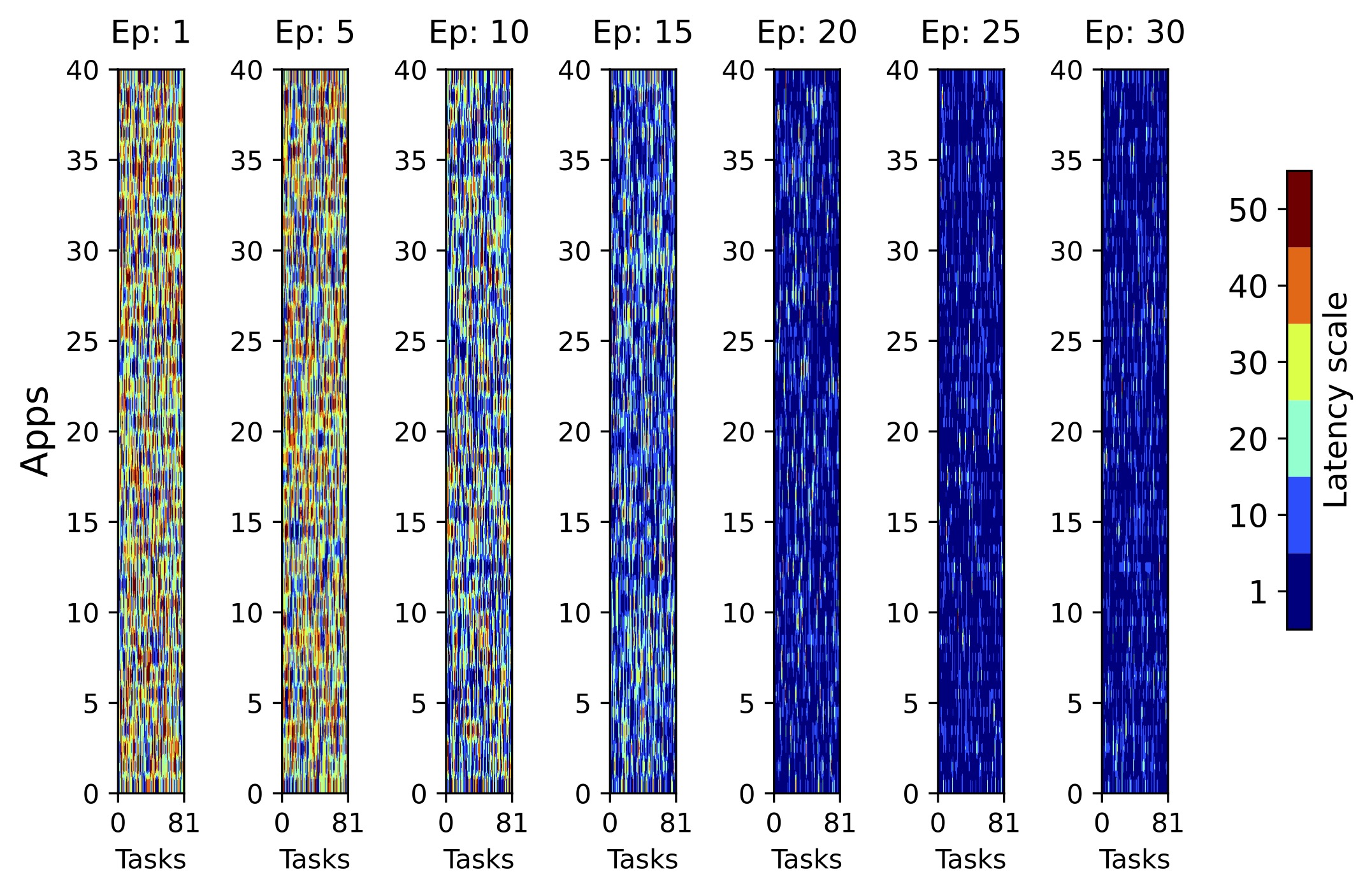}%
\label{fig:evolA}}
\hfil
\subfloat[]{\includegraphics[width=1.5in]{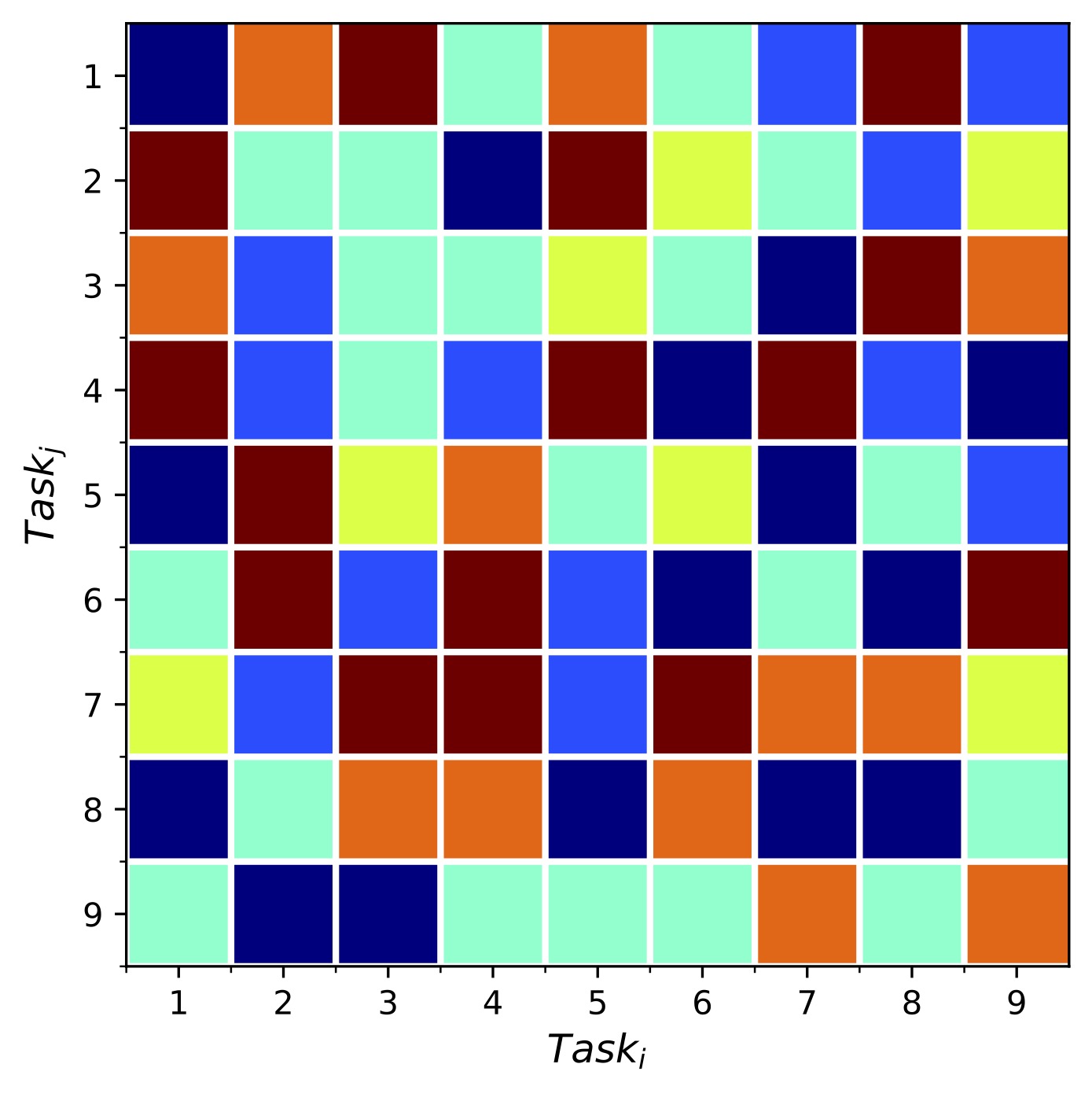}%
\label{fig:app0_1}}
\subfloat[]{\includegraphics[width=1.5in]{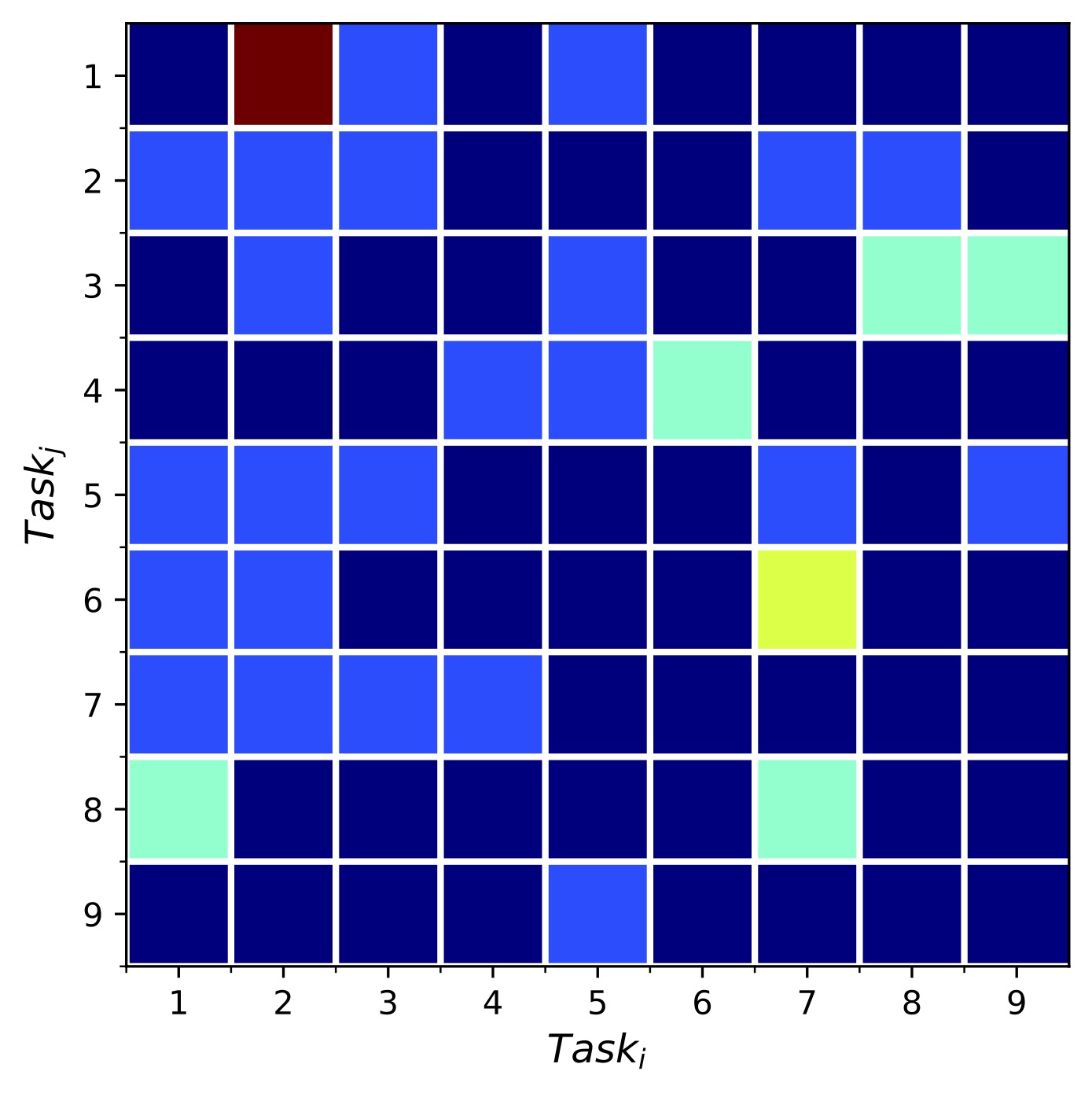}
\label{fig:app0_2}}
\subfloat[]{\includegraphics[width=2.in]{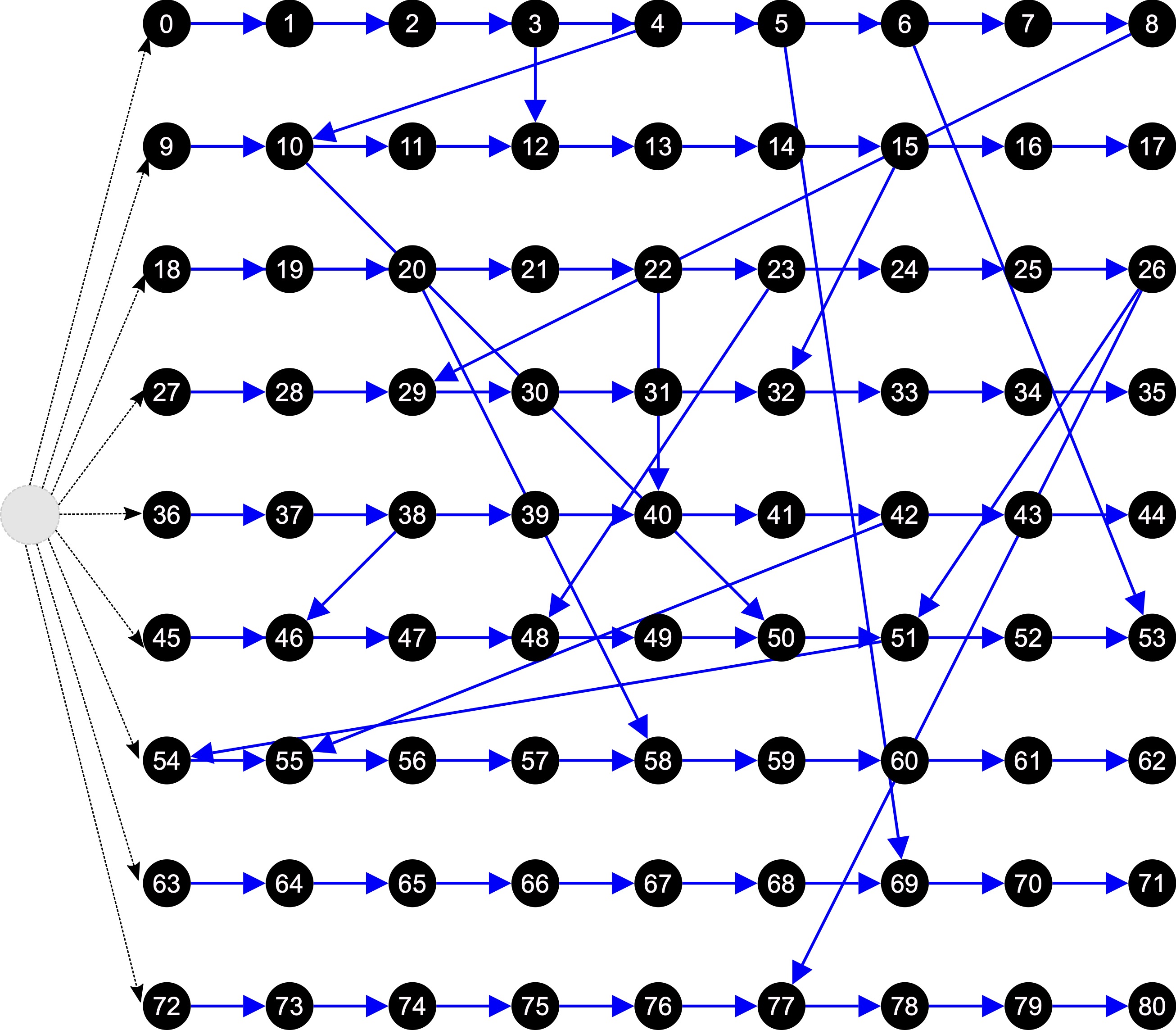}
\label{fig:appDAGone}}

\caption{\blue{Time-dependent allocation analysis for a configuration of 1000 devices and 81 services. (a) Heat map showing the evolution of latency-based device selection across different training episodes. (b) and (c) Heat maps of latency values for the first application in episodes 1 and 30, respectively. (d) DAG of service dependencies for the first application in this scenario.}}
\label{fig:evolucionlatencia}
\end{figure*}

The training evolution is consistent with both objectives. Figure~\ref{fig:evolucionlatencia} illustrates the evolution of device selection for each service during several training episodes. Thus, the main Figure~\ref{fig:evolA} shows a snapshot of the latency values of the assigned devices for each service in 7 different episodes: 1, 5, 10, 15, 20, 25 and 30. Each episode is a bar with a shape of 40 rows and 81 columns. Each row corresponds to a training environment, indicating that 40 applications are used in each episode. In this experiment, each environment comprises 1000 devices and 81 services ($9\times9$ services) by application. Hence, each column corresponds to one service. Each cell within the bar chart is colour-coded based on the service latency that has its assigned device. Thus, each row in the bar chart represents a placement scenario of the application along with the corresponding latency assignments. As the training progresses, the colour spectrum tends towards blue shades, indicating that the learning process assigns devices with lower latency. We provide a higher-resolution representation of a particular application in two episodes: Figure~\ref{fig:app0_1} in episode 1 and Figure~\ref{fig:app0_2} in episode 30. In both figures, we represent the application under a $9\times9$ shape instead of $1\times81$ as the main figure. It is worth noting that the algorithm considers the dependencies of each application to improve the allocation. We include the DAG of dependencies represented in Figure~\ref{fig:appDAGone} of this specific application. The evolution of the second objective, the cost of the placement, exhibits similar behaviour.

\subsubsection{Experiment II - Comparison with another multi-objective algorithm}

In this experiment, we compare the runtime and results with a GA and an NSGA-II with our DRL-based proposal.  GA and NSGA-II algorithms provide solutions for a specific environment configuration, requiring recomputation if there are changes in the environment. In contrast, our proposed DRL entails an exhaustive training period where the model learns from various environmental scenarios, simplifying the process of obtaining a solution for a new change.

The methodology for training the DRL model involves creating three datasets: a training dataset, a test dataset, and a validation dataset. The quality of the model is evaluated periodically during the training phase using the test set, allowing for the selection of the model that provides the best results. We use the validation dataset to define a new environment for these algorithms. Thus, we execute both GA algorithms on each specific scenario of the validation set.

For the runtime comparison, we compare the time required to train a single DRL model, which involves only a combination of weights between both objectives and the runtime to find a solution with NSGA-II and GA. In the case of the GA, we use the same similar combination of weights between objectives.






\begin{table*}
\caption{DRL model training time and runtime for NSGA-II and GA in hours for each experiment size.\label{table:comparativaTime}}
\centering
\begin{tabular}{ll|lll}
 \multicolumn{2}{c}{Scenario size} & \multicolumn{3}{|c}{Time (hours)}   \\\hline
Devices & Total Services & DRL Train &  NSGA-II & GA \\\hline
500        & 81   &  2:52:20  &   0:17:31  &  0:33:51  \\
500        & 144  &  8:18:47  &   0:31:00  &  0:56:44   \\\hline

1000        & 81   & 2:59:16 &  0:35:31  &  1:15:10   \\
1000       & 144  & 8:25:36 &  1:00:18  &  5:26:15   \\\hline

1500        & 81   &  3:03:53    & 0:52:18    &  1:51:14 \\
1500       & 144  &  8:35:15    &  1:33:07   &   7:27:39 \\\hline

2000        & 81   & 3:08:45     & 1:15:16  & 2:32:42 \\
2000       & 144  & 8:42:05     & \blue{1:51:34}  & ---  \\\hline 

\end{tabular}
\end{table*}

Table~\ref{table:comparativaTime}  presents results for various scenario sizes, ranging from 500 to 2000 devices, and two combinations of services 9 and 12 per row (or 81 and 144 services, respectively). The DRL-train column shows the runtime for the training phase for 150 episodes. The other two columns show the runtime taken by NSGA-II and GA for 200 generations to find a suitable solution.

The placement of an application within a model of 1000 devices and 81 services in total results in the following execution times: 63ms for our trained DRL model, 35 minutes for NSGA-II, and 1 hour for the GA algorithm. In all the scenario sizes, this time is lower than 1 second in all the configurations. To simplify the table, we do not include these values. 

From the values in the table, we observe that as the number of services in the application increases from 81 to 144, the execution time for each model significantly increases. This is attributed to the many services and dependencies impacting the application's execution time. Moreover, the number of devices influences the runtime with an average increase of 5 and 20 minutes for DRL and NSGA-II, respectively, compared to the same number of services. This increase is even more significant in the GA.

We compute the training time of the DRL model for the totality of episodes (150 in this experiment), but the convergence of results in many cases requires the range of 60-100 epochs. It is worth mentioning that hyperparameters of the model influence the execution time. For example,  reducing the number of layers in both Actors and Critics from 5 to 2 means a reduction of 2 hours for the case of 1500 devices and 144 services.

Another important aspect of this experiment is to evaluate the suitability of the solutions generated by the algorithms. We know the optimal placements since we have designed the experiment with devices that satisfy both latency and cost objectives. Furthermore, without any capacity limitation, the simplest and most optimal solution is to deploy all services on one of these devices. By using these optimal solutions, we can analyze how closely the solutions generated by these algorithms align with the optimal ones. 

The existence of a device with these characteristics causes the Pareto set to lean towards the latency and cost values presented by that device. This behaviour is what we are going to analyze with the NSGA-II algorithm and our DRL model. \blue{In our case, we train the initial model by balancing both objectives at 50\%, and then we transfer the state to different weight combinations (0-1, 0.25-0.75, 0.75-0.25, 1-0) to obtain a total of 5 solutions or 5 models, as illustrated in the space decomposition in Figure~\ref{fig:espacioescalar}.}

\begin{figure*}[!t]
\centering
\includegraphics[width=5.0in]{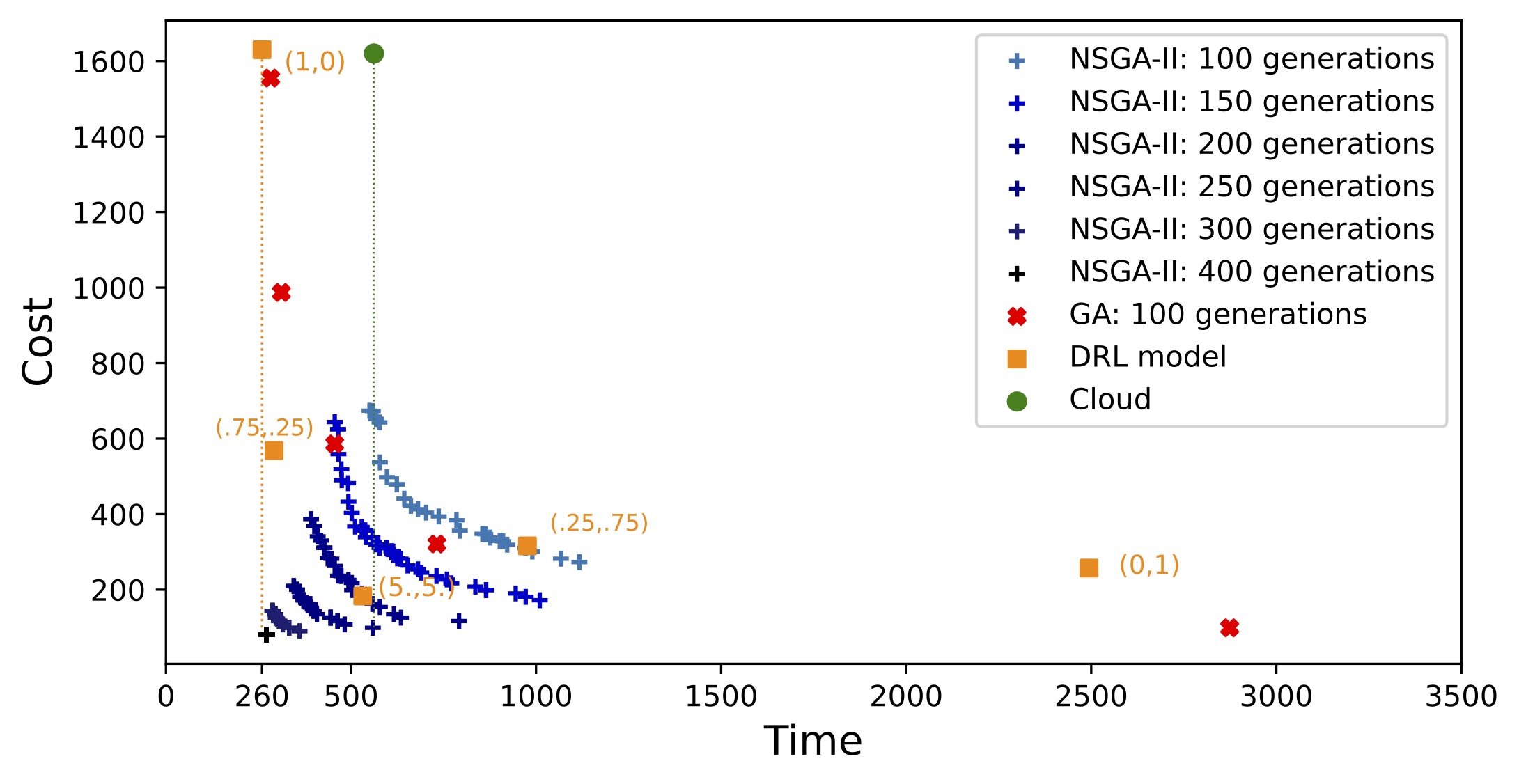}%
\caption{Solutions of the NSGA-II, GA and our DRL model for a scenario configuration of 1000 devices and 81 services per application.}
\label{fig:compa}
\end{figure*}

Figure~\ref{fig:compa} presents an analysis of the solutions obtained by our DRL model and by NSGA-II with a different number of generations. The green circle represents the values of the cloud devices. The five solutions obtained by our proposal are shown in orange, with the weights used listed alongside each solution. We also include the GA solutions, although we have not executed as many generations, as we consider the multi-objective proposal of the NSGA-II to be more representative for analyzing the spectrum of solutions and their dominance. 

Setting weights to guide rewards toward goals is not straightforward. For example, the best solution in terms of response time is the configuration with 100\% time and 0\% cost, while the best solution in terms of cost is the combination with 50\% time and 50\% cost.
NSGA-II reaches some of the solutions of the DRL model after 300 generations. The runtime for 300 generations is 2 hours and 5 minutes, while that for 400 generations is 2 hours and 56 minutes. The multiple waves of the Pareto front trend the convergence towards the ideal solution, which coincides with the values of the greedy strategies explained in the first experiment.
The GA solutions represent an approximate front end to those obtained by the DRL model. The runtime for each combination of weights in the GA is approximately 5 hours (25 hours for the five GA solutions).

\section{Conclusions and future work}

In this study, we propose a multi-objective deep reinforcement learning approach for service placement that considers the dependencies between services when making placement decisions.
Application placement involves placing independent and interrelated services that constitute the application. Therefore, we define a model capable of handling these dependencies to choose the service and subsequently, find the best device to deploy it.
In this approach, we define the model with a Graph Isomorphism Network (GIN) and two Proximal Policy Optimization agents (PPO) applied in two agent-critical models. Thus, the dependencies and features of the services feed the GIN model and an actor-critical model to choose a service. The second agent-critic uses the selected service, the device features and current service placements to choose a device.
This design has yielded promising results that align with the expected values obtained in our experiments, where we have analyzed the evolution of the learning model and the spectrum of solutions against a single-objective and a multi-objective evolutionary algorithm such as NSGA-II.

Our proposal addresses the multi-objective problem by redirecting the learning reward with a weighting combination. Thus, we decompose the problem space into a mix of weights, and in turn, we reduce the training cost by transferring weights between the different models, one for each combination. The experimental results show that our approach produces results consistent with those obtained using different genetic algorithms. This type of solution provides mechanisms that can offer real-time solutions once trained. Our trained model takes less than a second to deploy an application, whereas using an NSGA-II or GA model can take minutes or even hours. However, the feasibility of converting a DRL model into a multi-objective depends on how independent each objective is in its representation within the neural model. In our approach, the same neural model combines both goals and may tend to optimize any of them, so there is no guarantee that all solutions will be non-dominated solutions in the combination of weights. \blue{Despite the rapid acquisition of solutions, a multi-objective genetic algorithm offers a more diverse set of solutions than our approach. Exploring solutions for multi-objective optimization problems using DRL-based methods presents interesting challenges for future research.}

We find different research lines for future work promising, particularly those aimed at enhancing the simplification of defining rewards for multi-objective functions with multiple constraints. \blue{It could be interesting to define dynamic reward adjustment mechanisms} during the training phase towards other objectives, even if some goals have been achieved previously and evaluate whether a unique model of this type satisfies the requirements of an infrastructure provider. \blue{Additionally, it will be valuable to analyze the solution front considering constraints and a larger set of objectives. }

\bmhead{Acknowledgments}

Funding: MCIN/AEI/10.13039/501100011033 [grant number PID2021-128071OB-I00] and by FEDER, UE.

\bibliography{sn-bibliography}

\end{document}